\documentclass[12pt,preprint]{aastex}

\usepackage{url}

\newcommand{\CaII}{\ion{Ca}{2}}
\newcommand{\CaIIH}{\ion{Ca}{2}~H}
\newcommand{\kms}{km$\,$s$^{-1}$}
\newcommand{\mss}{m~s$^{-2}$}

\newcommand{\ha}{H$\alpha$}
\newcommand{\sm}{$\sim$}
\newcommand{\hinode}{\textit{Hinode}}

\begin{document}

\title{CHROMOSPHERIC RAPID BLUESHIFTED EXCURSIONS OBSERVED WITH IBIS AND THEIR ASSOCIATION WITH PHOTOSPHERIC MAGNETIC FIELD EVOLUTION}

\author{Na Deng\altaffilmark{1,2}, Xin Chen\altaffilmark{1}, Chang Liu\altaffilmark{1,2}, Ju Jing\altaffilmark{1,2}, Alexandra Tritschler\altaffilmark{3},\\Kevin P. Reardon\altaffilmark{3,4}, Derek A. Lamb\altaffilmark{5}, Craig E. Deforest\altaffilmark{5}, Carsten Denker\altaffilmark{6}, Shuo Wang\altaffilmark{1}, Rui Liu\altaffilmark{7}, and Haimin Wang\altaffilmark{1,2}}
\affil{$^1$~Space Weather Research Laboratory, New Jersey Institute of Technology, University Heights, Newark, NJ 07102-1982, USA; na.deng@njit.edu}
\affil{$^2$~Big Bear Solar Observatory, New Jersey Institute of Technology, Big Bear City, CA 92314-9672, USA}
\affil{$^3$~National Solar Observatory, Sacramento Peak, Sunspot, NM 88349-0062, USA;}
\affil{$^4$~INAF-Osservatorio Astrofisico di Arcetri, Largo E. Fermi 5, 50125 Florence, Italy;}
\affil{$^5$~Southwest Research Institute, 1050 Walnut Street, Suite 300, Boulder, CO 80302-5142, USA;}
\affil{$^6$~Leibniz-Institut f\"ur Astrophysik Potsdam, An der Sternwarte 16, 14482 Potsdam, Germany;}
\affil{$^7$~Department of Geophysics and Planetary Sciences, University of Science and Technology of China, Hefei 230026, China}

\begin{abstract}
Chromospheric rapid blueshifted excursions (RBEs) are suggested to be the disk counterparts of type II spicules at the limb and believed to contribute to the coronal heating process. Previous identification of RBEs was mainly based on feature detection using Dopplergrams. In this paper, we study RBEs on 2011 October 21 in a very quiet region at the disk center, which were observed with the high-cadence imaging spectroscopy of the \CaII\ 8542~\AA\ line from the Interferometric Bidimensional Spectrometer (IBIS). By using an automatic spectral analysis algorithm, a total of 98 RBEs are identified during a 11 minute period. Most of these RBEs have either a round or elongated shape, with an average area of 1.2 arcsec$^2$. The detailed temporal evolution of spectra from IBIS makes possible a quantitative determination of the velocity (\sm16~\kms) and acceleration (\sm400~\mss) of \CaII\ 8542 RBEs, and reveal an additional deceleration (\sm$-$160~\mss) phase that usually follows the initial acceleration. In addition, we also investigate the association of RBEs with the concomitant photospheric magnetic field evolution, using coordinated high-resolution and high-sensitivity magnetograms made by \hinode. Clear examples are found where RBEs appear to be associated with the preceding magnetic flux emergence and/or the subsequent flux cancellation. However, a further analysis with the aid of the Southwest Automatic Magnetic Identification Suite does not yield a significant statistical association between these RBEs and magnetic field evolution. We discuss the implications of our results in the context of understanding the driving mechanism of RBEs.
\end{abstract}

\keywords{Sun: chromosphere -- line: profiles -- Sun: activity -- Sun: photosphere -- Sun: magnetic fields}

\section{INTRODUCTION}\label{sec:introduction}
The solar chromosphere, even in the quiet Sun region, is well known to exhibit small-scale jets in the form of spicules \citep[e.g.,][]{beckers1968}. Using high-resolution and high-cadence solar-limb observations in \CaIIH\ (3968~\AA) from the Solar Optical Telescope \citep[SOT;][]{Tsuneta+etal2008SoPh..249..167T} on board \hinode, \citet{DePontieu+etal2007PASJ...59S.655D} classified two types of spicules. The differences between them were statistically investigated by \citet{Pereira+etal2012ApJ...759...18P} using \hinode/SOT data, and further revealed by \citet{2014ApJ...792L..15P} using spectra and filtergrams from the recently launched \textit{Interface Region Imaging Spectrograph}. These studies suggest that Type I spicules have a speed of 15--40~\kms\ and remain visible in chromospheric passbands during their lifetime. In contrast, type II spicules are of particular importance because of their higher speed (30--110~\kms) and rapid heating to higher transition region temperatures. It was also found that type II spicules undergo two other kinds of motions in addition to the field-aligned flows: transversal swaying motion and torsional motion \citep{DePontieu+etal2007Sci...318.1574D, DePontieu+etal2012ApJ...752L..12D}.

Since type II spicules were discovered at the solar limb, identifying their disk counterparts is important. Previous disk observations of the quiet Sun by the Big Bear Solar Observatory using the Littrow Spectrograph identified conspicuous absorption features in the blue wing of H$\alpha$, which are located around concentrations of enhanced magnetic activity \citep{WangH+etal1998SoPh..178...55W,Chae+etal1998ApJ...504L.123C,chae99b,LeeCY+etal2000ApJ...545.1124L}. Different from classical spicules that show absorption in both wings, these short-lived features exhibit absorption in the blue wing only and were thus named upflow events (UFEs). The recent search for the disk counterparts of type II spicules has been carried out with high-quality ground-based spectral imaging observations. Using the Interferometric Bidimensional Spectrometer \citep[IBIS;][]{Cavallini2006SoPh..236..415C,Reardon+Cavallini2008A&A...481..897R} at the National Solar Observatory (NSO), \citet{Langangen+etal2008ApJ...679L.167L} found disk events showing a blueward Doppler shift only in the blue wing of \CaII~8542~\AA\ spectra and named them rapid blueshifted excursions (RBEs), which most probably belong to the \ha\ UFEs discovered earlier. These RBEs and type II spicules share some similar properties (e.g., rapid motion and location), which suggests their common origin. In another study, using the CRisp Imaging SpectroPolarimeter (CRISP) \citet{Rouppe+etal2009ApJ...705..272R} resolved RBEs as thin, jet-like features possessing a Doppler velocity of order 20~\kms\ and 50~km~s$^{-1}$ in the \CaII\ 8542~\AA\ and \ha\ lines, respectively. The plasma is observed to be accelerated and rapidly heated along the jet resembling type II spicules. Very recently, \citet{Sekse+etal2012ApJ...752..108S,Sekse+etal2013ApJ...764..164S} applied an automatic detection routine to CRISP Dopplergrams and found the occurrence rate and the three different types of motions of RBEs comparable to those of type II spicules. The authors showed a continuous motion in both \CaII\ 8542~\AA\ and \ha\ spectral data obtained simultaneously, except that \CaII\ 8542 RBEs are located closer to the point of origin and appear earlier than their \ha\ counterpart. They also found that RBEs in disk-center quiet-Sun regions have lower Doppler velocities than those in coronal holes.

Although significant progress on the study of type II spicules and RBEs has been made indicating that they are most probably manifestations of the same phenomenon, the exact physical origin of this jet activity is still unknown. The main driving mechanisms of UFEs/RBEs proposed thus far include small-scale magnetic reconnection \citep[e.g.,][]{WangH+etal1998SoPh..178...55W,Chae+etal1998ApJ...504L.123C,Sterling+etal2010ApJ...714L...1S,Moore+etal2011ApJ...731L..18M}, plasma squeezing due to small-scale flux emergence \citep{MartinezSykora+etal2011ApJ...736....9M,MartinezSykora+etal2013ApJ...771...66M}, and Alfv\'en waves \citep[e.g.,][]{Hollweg+etal1982SoPh...75...35H, DePontieu+Haerendel1998A&A...338..729D,DePontieu+etal2012ApJ...752L..12D}. As small-scale jets are presumed to play an important role in providing the upward flux of energy and momentum, the study of RBEs will advance the current understanding of the coronal heating and solar wind \citep{tsiropoula12}. It should also be noted that the physical nature of spicule-like, rapidly changing chromospheric structures are still under active discussion \citep[e.g.,][]{2012ApJ...755L..11J,2014ApJ...785..109L}.

Previous identification of RBEs has mainly depended on analyzing the spicular features seen in Dopplergrams \citep[e.g.,][]{Sekse+etal2012ApJ...752..108S}. In contrast, an alternative approach on a pure spectroscopic basis \citep{Langangen+etal2008ApJ...679L.167L} has not been fully explored. Also importantly, little attention has been paid to the photospheric magnetic field evolution associated with RBEs, which could provide constraints on the various suggested driving mechanisms \citep{Rouppe+etal2009ApJ...705..272R}. In this paper, we present an analysis of IBIS observations of \CaII\ 8542 RBEs in a quiet-Sun region, using an automatic detection algorithm based on spectral characteristics. We also study the concomitant evolution of magnetic elements on the photosphere using the coordinated high-resolution and high-sensitivity magnetograms made by \hinode. In addition, an exploratory examination of the statistical association of RBEs with magnetic field dynamics is made. The plan of this paper is as follows: in Section~\ref{sec:observation}, we describe the data sets and
reduction procedure. In Section~\ref{sec:results}, we show the main results
of data analysis and discuss their implications. Section~\ref{sec:summary} summarizes and discusses major findings.

\section{OBSERVATIONS AND DATA PROCESSING}\label{sec:observation}
We conducted an observing campaign with the IBIS instrument and \hinode\ telescope from 2011 October 17 to 23 under the \hinode\ Operation Plan No. 203\footnote{\url{http://www.isas.jaxa.jp/home/solar/hinode_op/hop.php?hop=0203}}. IBIS observed chromospheric RBEs in \ha\ and \CaII\ 8542~\AA\ with the 76~cm Dunn Solar Telescope at NSO/Sacramento Peak \citep{Zirker1998SoPh..182....1Z} equipped with adaptive optics \citep{rimmele+etal2004, Rimmele+Marino2011LRSP....8....2R}. \hinode/SOT targeted at the same region as IBIS, and recorded with its Narrowband Filter Imager (NFI) high-spatial (0$\farcs$32) and high-temporal (64~s) resolution photospheric magnetograms in the line of sight (LOS) using the Na\,\textsc{i}\,D$_1$ (5896~\AA) line.

This work presents the continuous IBIS observation of a quiet-Sun region taken on 2011 October 21 from 16:08:02 to 16:19:31~UT (e.g., Figure~\ref{f1}(a)), when there was a relatively stable, moderate to good seeing condition. The region is very close to the disk center at solar coordinates ($x$, $y$) $\approx$ (32\arcsec, $-$50\arcsec) ($\mu = 0.998$).  The disk-center observation minimizes the projection and LOS effects providing a better view of the root of the chromospheric RBEs. We concentrate on the time series of \CaII\ 8542~\AA\ data rather than \ha, because (1) RBEs are seen to be located closer to the point of origin in \CaII\ 8542 \citep{Sekse+etal2012ApJ...752..108S, Sekse+etal2013ApJ...764..164S}, and (2) a better image stability can be achieved in \CaII\ 8542 due to its longer wavelength \citep{Roddier1999aoa..book....9R}.

IBIS has two synchronized channels, which were set as described below when observing with the \CaII\ 8542~\AA\ line. The narrowband channel (with 0.044~\AA\ FWHM around 8542~\AA) used two tunable Fabry-P\'erot interferometers (FPI) to take bidimensional spectral images from 8540.2 to 8543.5~\AA~(i.e., from $-$1.9~\AA\ to $+$1.4~\AA) across the Ca line center (8542.1~\AA) at 34 unevenly spaced steps. A prefilter centered around 8542.1~\AA\ with an FWHM of 4.6~\AA\ was mounted to isolate one of the periodic transmission peaks of the FPIs. Meanwhile, the broadband (\sm100~\AA) channel took white-light reference images at 8300~\AA\ for calibration and post-facto image processing. Each full spectral scan took about 5.85~s, at a frame rate of \sm6 frames s$^{-1}$ and with an exposure time of 30~ms frame$^{-1}$. A total of 118 spectral scans were obtained between 16:08:02 and 16:19:31~UT on 2011 October 21. The image scale is about 0$\farcs$2~pixel$^{-1}$ after 2~$\times$~2 binning in post processing (for a more manageable data size and an enhanced signal-to-noise ratio), and the round field of view (FOV) has a diameter of 95\arcsec.

To process IBIS data, we first performed the standard calibration procedures, which include dark and flat-field correction, alignment and destretch of images in each spectral scan using broadband white-light images as a reference, and correction of blue shifts across the FOV due to the collimated mount of the IBIS FPIs. To alleviate the impact of the prefilter transmission curve $T$ that is superimposed on all the observed spectral profiles $I$, we first divide an averaged line profile of the disk-center flat field by the corresponding profile segment cropped from the atlas of the disk center. The atlas of the solar flux spectrum is produced by the Kitt Peak Fourier transform spectrometer \citep{Neckel+Labs1984SoPh...90..205N, Neckel1999SoPh..184..421N}. The resulted curve was then smoothed with a high-degree polynomial fit, and this approximated $T$ was used to correct the observed spectra by $I/T$.

The corresponding \hinode/NFI magnetogram data (e.g., Figure~\ref{f1}(b)) span from 15:00 to 17:10~UT, which well covers the photospheric magnetic field evolution related to the IBIS observation. To calibrate the NFI data and reduce the noise, we followed the procedures of \citet{Lamb+etal2010ApJ...720.1405L}. First the spatially variable offset in the images due to the two slightly different NFI detectors was removed based on a linear fit to the median value of each column. The data were then processed using the standard \hinode/SOT calibration routine \verb+fg_prep.pro+ in the solar software, which also cleaned the spikes produced by cosmic rays. The entire image time sequence was subsequently aligned with sub-pixel precision. Finally, a temporal smoothing (with a 2 minute FWHM Gaussian weighting) and a spatial smoothing (using a truncated 3$\times$3 Gaussian kernel with a 2 pixel FWHM) were applied. Compared to the raw data, the noise of the processed magnetograms is reduced by 62\% to $\sigma=5.5\pm0.2$~G.

We tried to achieve the best possible alignment between chromospheric and photospheric images, which is non-trivial for the quiet-Sun observation. The time sequence of 118 spectral scans of IBIS was first self-aligned, after which a co-registration with the processed \hinode/NFI magnetic field images was made by matching the network field structure. Specifically, we used blue-wing intensity images at $\Delta\lambda=1.6$~\AA\ from the \CaII\ 8542 line center, which is sensitive to the presence of magnetic features \citep[e.g.,][]{Vecchio+etal2007A&A...461L...1V}. The accuracy of the alignment is estimated to be $\lesssim$1\arcsec.

Quantitatively identifying and tracking small-scale magnetic activities are a challenging task. Here we implement the Southwest Automatic Magnetic Identification Suite (SWAMIS\footnote{\url{http://www.boulder.swri.edu/swamis}}), which is a demonstrated technique for automatic magnetic feature detection and tracking using advanced methods (such as dual-threshold scheme for feature discrimination and dual-maximum overlap
criterion for feature association) and has been extensively used on \hinode/NFI data. For the detailed working procedure of SWAMIS, we refer readers to the literature \citep{DeForest+etal2007ApJ...666..576D, Parnell+etal2009ApJ...698...75P, Lamb+etal2008ApJ...674..520L, Lamb+etal2010ApJ...720.1405L, Lamb+etal2013ApJ...774..127L}. The application of SWAMIS on this particular \hinode/NFI data set is also elaborated in a separate work \citep{chen14}.

\section{RESULTS AND ANALYSIS}\label{sec:results}
We first present detection and characterization of chromospheric RBEs (Section~\ref{sec:RBE}), then proceed to examine some examples of RBEs with specific evolution of photospheric magnetic field (Section~\ref{sec:CanEmg}). At last we explore the statistical association of RBEs with magnetic field dynamics (Section~\ref{sec:RBEwithCanEmg}).

\subsection{Detection and Properties of Disk-center Quiet-Sun \CaII\ 8542 RBEs}\label{sec:RBE}
The line profiles of UFEs found by \citet{LeeCY+etal2000ApJ...545.1124L} show absorption in the blue wing only, whereas the red wing could display either little deviation (e.g., the red profile in Figure~\ref{f3}(m)) or obvious intensity enhancement (e.g., the blue profile in Figure~\ref{f3}(m)) when compared to the quiet-Sun profile. The latter case is equivalent to that the whole line shifts toward the blue. Strictly speaking, RBEs conform to the former kind of UFEs as they are characterized by a Doppler shift that only appears in the blue wing \citep[e.g.,][]{Langangen+etal2008ApJ...679L.167L}. Nevertheless, as there may not be a fundamental physical difference between the two kinds of UFEs \citep{LeeCY+etal2000ApJ...545.1124L}, we do not distinguish explicitly between UFEs and RBEs in this study. Here we develop an algorithm for detecting this type of events based on the spectral characteristics of features particularly in the blue wing. Specifically, the automated procedure examines the spectral line profile at each pixel to evaluate whether it has strong absorption in the blue wing only. In practice, the criterion is considered met when compared to the reference profile (the averaged profile over the entire FOV), the additional absorption in the blue wing is greater than 18\% ($\sim$3$\delta$; here $\delta$ is the relative standard deviation of the image intensity) and that in the red wing is less than 6\% ($\sim$1$\delta$), at $\Delta\lambda=-0.4$~\AA\ and 0.4~\AA\, respectively, from the \CaII\ 8542 line center. At each instance all the qualified, connected pixels are clustered together to form RBE features. Since RBEs may exhibit motion, the overlapping area of an RBE feature in at least three consecutive spectral scans (\sm18~s) is defined as the region of this particular RBE. Finally, RBE regions that incorporate fewer than six pixels are removed, and directly neighboring RBE regions are merged. In this fashion, within our 95\arcsec\ diameter circular FOV we identified 98 \CaII\ 8542 RBEs distributed over the 11 minute time period from 2011 October 21 16:08:02~UT to 16:19:31~UT. Since our method takes full advantage of the spectral information in each pixel, it provides an appropriate and effective RBE detection by avoiding false and/or missed identifications, which might be an issue when only Doppler images are used.

\begin{figure}[t]
  \epsscale{1.}
  \plotone{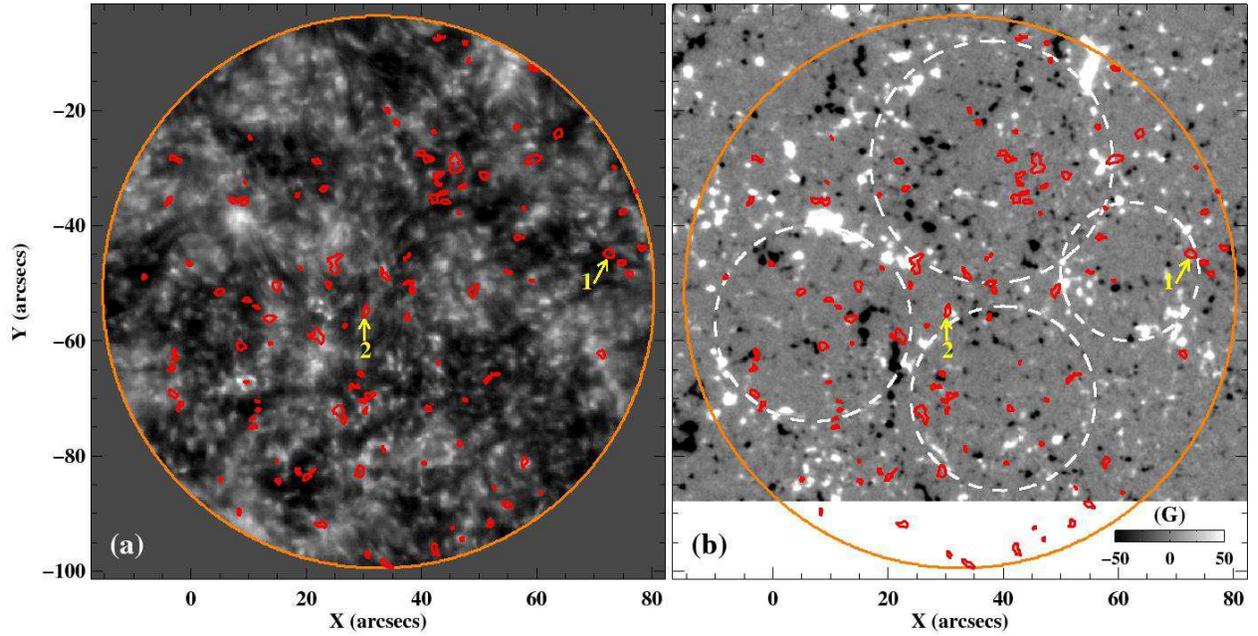}
  \caption{IBIS \CaII\ 8542~\AA\ (a) and \hinode/NFI LOS magnetic field (b) observations of a very quiet region at the disk center at 16:10 UT on 2011 October 21. The red contours outline the detected 98 RBEs during 16:08:02--16:19:31~UT, 88 of which are in the FOV of \hinode\ observations (note the magnetogram is scaled from $-$50 to 50~G). The event 1 and 2, as pointed to by arrows, are presented in Figures~\ref{f2} and \ref{f3}, respectively. The dashed white circles in (b) roughly delineate the supergranular cells. The orange circle indicates the FOV of IBIS.}
  \label{f1}
\end{figure}

In Figure~\ref{f1}, we use red contours to outline the detected 98 RBEs, 88 of which are in the FOV of \hinode\ observations. The RBEs are seen to be located mainly around the boundary of the supergranular cells (depicted by the dashed lines in Figure~\ref{f1}(b)) with some residing inside the cells, as also observed by \citet{LeeCY+etal2000ApJ...545.1124L}. Based on a visual inspection, the shape of these disk-center quiet-Sun RBEs is mostly round or elongated, similar to the result of \citet{Langangen+etal2008ApJ...679L.167L} for a disk-center region. It is, however, noticeable that in all the previous RBE studies, the target region, either in the quiet-Sun or coronal holes, had obvious magnetic concentrations; accordingly, RBEs were mainly elongated and found in the inner edge of the conspicuous chromospheric rosette structures \citep{Langangen+etal2008ApJ...679L.167L,Rouppe+etal2009ApJ...705..272R,Sekse+etal2012ApJ...752..108S,Sekse+etal2013ApJ...764..164S,Sekse+etal2013ApJ...769...44S}. In contrast, we are looking at a very quiet region as shown by the highly sensitive \hinode\ LOS magnetogram; no well-formed fibril or rosette structure is discernible in the \CaII\ 8542 line center image (Figure~\ref{f1}(a)).

In order to further characterize the RBE occurrence, we create $\lambda$-t plots by stacking the averaged line profile $I$($\lambda$) within an RBE region over time. These kind of $\lambda$-t plots can also be viewed as $\upsilon_\mathrm{Doppler}$-t plots, where the Doppler velocity $\upsilon_\mathrm{Doppler}$ is related to the corresponding wavelength $\lambda_\mathrm{c}$ following

\begin{equation}
\upsilon_\mathrm{Doppler}=\frac{c}{\lambda_0}(\lambda_\mathrm{c}-\lambda_0) \ ,
\end{equation}

\noindent where $c$ is the speed of light and $\lambda_0$ is the wavelength of the resting line center (i.e., 8542.1~\AA). For a more precise estimation of $\upsilon_\mathrm{Doppler}$ of the blueshifted spectral component, we follow the method introduced by \citet{Rouppe+etal2009ApJ...705..272R}:

\begin{equation}
\upsilon_\mathrm{Doppler}=\frac{c}{\lambda_0}\frac{\int_{\lambda_\mathrm{min}}^{\lambda_0} (\lambda-\lambda_0)(I_\mathrm{ref}-I)~d\lambda}{\int_{\lambda_\mathrm{min}}^{\lambda_0} (I_\mathrm{ref}-I)~d\lambda} \ ,
\end{equation}

\noindent where $I$ is the intensity of the blueshifted profile and $I_\mathrm{ref}$ is the intensity of the reference profile. As the intensity fluctuations in the far blue wing can introduce large errors in estimating $\upsilon_\mathrm{Doppler}$, the integration is made from $\lambda_\mathrm{min}=8541.0$~\AA\ to $\lambda_0=8542.1$~\AA, which correspond to a $\upsilon_\mathrm{Doppler}$ ranging from $-$40~\kms\ to 0~\kms. In addition, the integration is made over the absorption portion of the spectrum (i.e., where $I < I_\mathrm{ref}$). The temporal evolution of spectra of two sample RBE events 1 and 2 (denoted in Figure~\ref{f1}) are drawn in Figures~\ref{f2}(l) and \ref{f3}(l), respectively, in which $\lambda_\mathrm{c}$ corresponding to the derived $\upsilon_\mathrm{Doppler}$ at each instance of time is marked by the plus sign, with the time of the maximum $\upsilon_\mathrm{Doppler}$ colored blue. These plus signs clearly portray that a typical RBE event can be visualized in the temporal evolution of spectra by an asymmetric V-shaped excursion in wavelength toward the blue wing. In Figures~\ref{f2}(m) and \ref{f3}(m), we also plot the blueshifted profiles, the reference profile (dashed black line), and their differences (most right group of lines) at some selected time instances (with the same color codes as those of the plus signs in Figures~\ref{f2}(l) and \ref{f3}(l)), in both the scales of $\Delta\lambda$ and $\upsilon_\mathrm{Doppler}$. These clearly demonstrate the characteristic, exclusive blue shifts of RBEs and their temporal evolution. We point out that around the time of maximum $\upsilon_\mathrm{Doppler}$, the average spectra of our defined RBE regions usually show a whole shift toward the blue, possibly indicating bulk motion. At other times, spectra with only an additional blue-wing component can be observed (see, e.g., the red profile in Figure~\ref{f3}(m)). 
 
The detailed temporal evolution of spectra from IBIS enables us to investigate the temporal evolution of $\upsilon_\mathrm{Doppler}$ and acceleration/deceleration of RBEs in an intuitive and quantitative fashion. We find that the maximum $\upsilon_\mathrm{Doppler}$ of our 98 RBEs ranges from 12.4 to 22.3~\kms\ with a mean of 16.1~\kms, and that the lifetime of RBEs in Ca~{\sc ii} 8542~\AA\ ranges from 17.6 to 99.4~s with a mean of 39~s. These results agree with previous studies using CRISP data \citep[e.g.,][]{Rouppe+etal2009ApJ...705..272R, Sekse+etal2012ApJ...752..108S,Sekse+etal2013ApJ...764..164S}, despite that the velocity is at the lower end. This could be because the spatial resolution of IBIS is lower than CRISP \citep{2013ApJ...764...69P}, and because our region of interest is very quiet, much different from the regions studied before. In addition, the area of these RBEs ranges from 0.28 to 5.5 arcsec$^2$ (with a mean of 1.2 arcsec$^2$) and is correlated with the RBE lifetime (with a correlation coefficient of 0.64), meaning that larger RBEs tend to live longer. Histograms of these physical properties are presented in Figure~\ref{f35}. Remarkably, we are also able to derive the properties of acceleration $a_f$ and deceleration $a_s$ of our RBEs in a straightforward way, by analyzing the slopes of edges of the V-shaped RBE feature in the $\upsilon_\mathrm{Doppler}$-t diagram. The study of tens of conspicuous cases gives an $a_f$ of \sm200--700 \mss\ with a mean of \sm400 \mss\ and a following $a_s$ ranging from \sm$-$90 to $-$240 \mss\ with a mean of \sm$-$160 \mss, the latter of which is generally smaller than the gravitational acceleration of \sm$-$274 \mss\ on the Sun. This suggests that our observed RBEs generally exhibit a fast acceleration followed by a slow deceleration toward blue. In some previous observations, evidence of acceleration of RBEs toward the end of their short appearance in chromospheric passbands was reported without a quantitative estimate \citep[e.g.,][]{DePontieu+etal2007PASJ...59S.655D, Rouppe+etal2009ApJ...705..272R}. Here we disclose, based on our high-cadence IBIS data, an additional deceleration phase that usually closely follows the initial acceleration. As a further comparison, the deceleration of jet-like dynamic fibrils in the chromosphere was measured earlier by fitting their paths in the distance-time plots \citep{Hansteen+etal2006ApJ...647L..73H, DePontieu+etal2007ApJ...655..624D}, the result of which is similar to ours.

It is worth mentioning that quite often the RBE signature seen in the temporal evolution of spectra seems to be related to the preceding and/or following redshift events, with a time gap of 1--2 minutes (see e.g., Figures~\ref{f2}(l) and \ref{f3}(l)). The overall shape appears sawtooth-like, reminiscent of the spectral signature of the consequence of acoustic shocks that propagate from the photosphere \citep[e.g.,][]{carlsson+stein1997,Vecchio+etal2009A&A...494..269V}. We cannot exclude the possibility that some of our detected RBEs, especially those located inside the cells, may include or be ``modulated'' by shocks. A quantitative analysis of these oscillations in relation to the RBE occurrence is not attempted in the present work. Nevertheless, many RBEs we found apparently occur at locations of magnetic structures as observed by \hinode\ (see further discussion below in Section~\ref{sec:CanEmg}), while previous studies suggested that shocks generally avoid magnetic elements \citep{Vecchio+etal2009A&A...494..269V}. Alternatively, the sawtooth-like spectral structure can be due to the coexistence of RBEs and rapid redshifted excursions (see Figure~2 of  \citealt{Sekse+etal2013ApJ...769...44S}), the latter of which are observed on-disk owing to the torsional or transversal motions possibly harbored by spicules.

\subsection{RBEs Associated with Photospheric Magnetic Flux Emergence and/or Cancellation}\label{sec:CanEmg}

\begin{figure}[t]
  \epsscale{1}
  \plotone{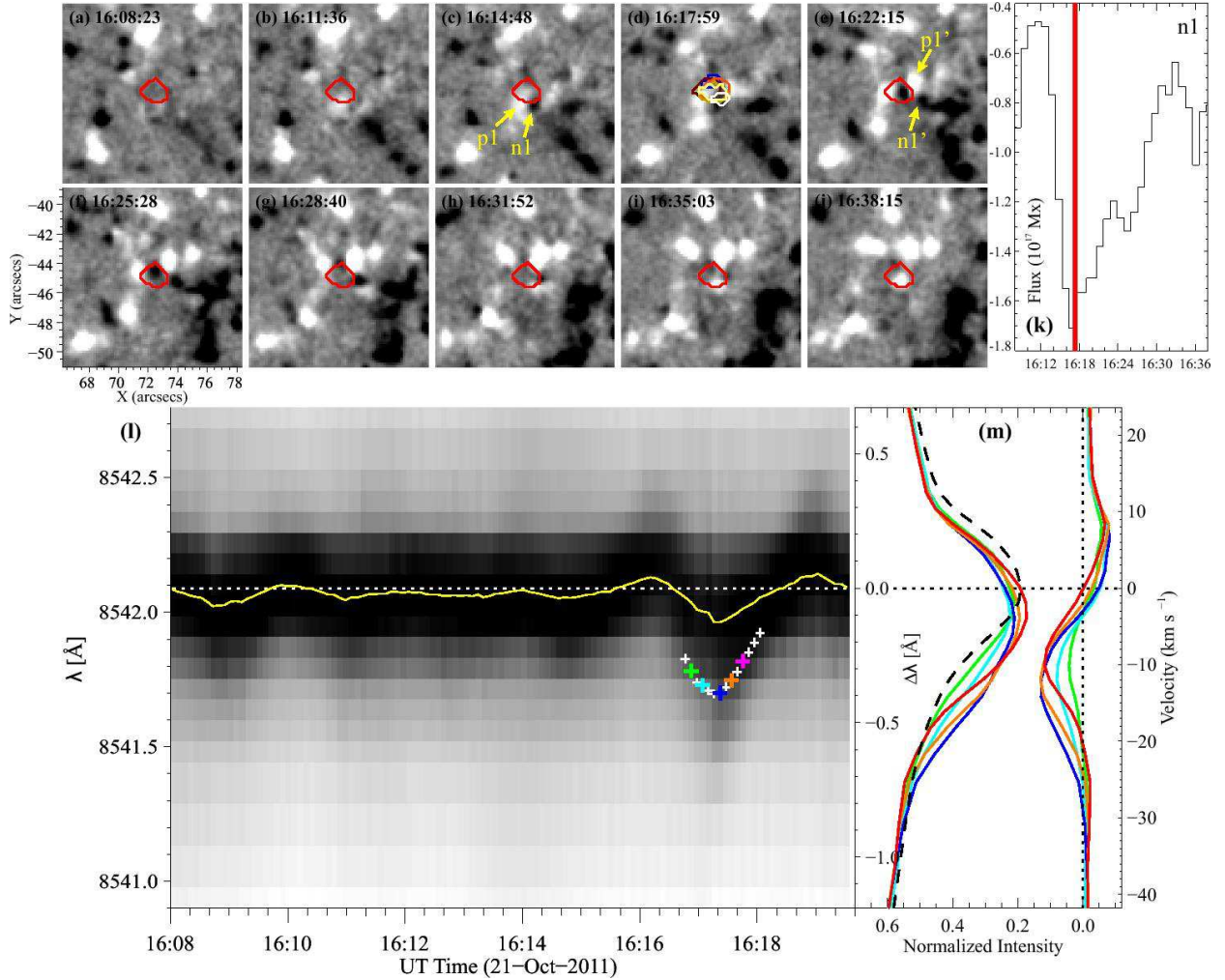}
  \caption{Evolution of the RBE event 1, as denoted in Figure~\ref{f1}. (a)--(j) show the emergence and the ensued cancellation of magnetic elements p1-n1 beneath the RBE. The \hinode/NFI LOS magnetograms are made saturated at $\pm$35~G to show the fine field structure. The color-coded contours in (d) outline the RBE feature at several instances based on the IBIS observation. Darker (brighter) colors represent earlier (later) times. The red contour in other panels depicts the defined region for this RBE (see Section~\ref{sec:RBE}). The temporal evolution of the magnetic flux of the negative polarity element n1 is plotted in (k), in which the red shade denotes the time interval of the RBE occurrence. (l) is a temporal evolution plot of the spectrum averaged over the RBE region. The horizontal white dotted line indicates the resting line center position. The solid yellow curve shows the temporal evolution of the measured line center position. The plus signs mark the derived Doppler velocity $\upsilon_\mathrm{Doppler}$ of the corresponding blueshifted line profiles, with the highest $\upsilon_\mathrm{Doppler}$ colored blue. In (m), some selected blueshifted profiles (with the same color codes as those of the pluses in (l)) and their differences (most right group of lines) between the reference profile (dashed black line) are plotted, with $y$ axes consistent with the panel (l). The reference profile is averaged over the entire FOV of IBIS.} \label{f2}
\end{figure}

\begin{figure}[t]
  \epsscale{1}
  \plotone{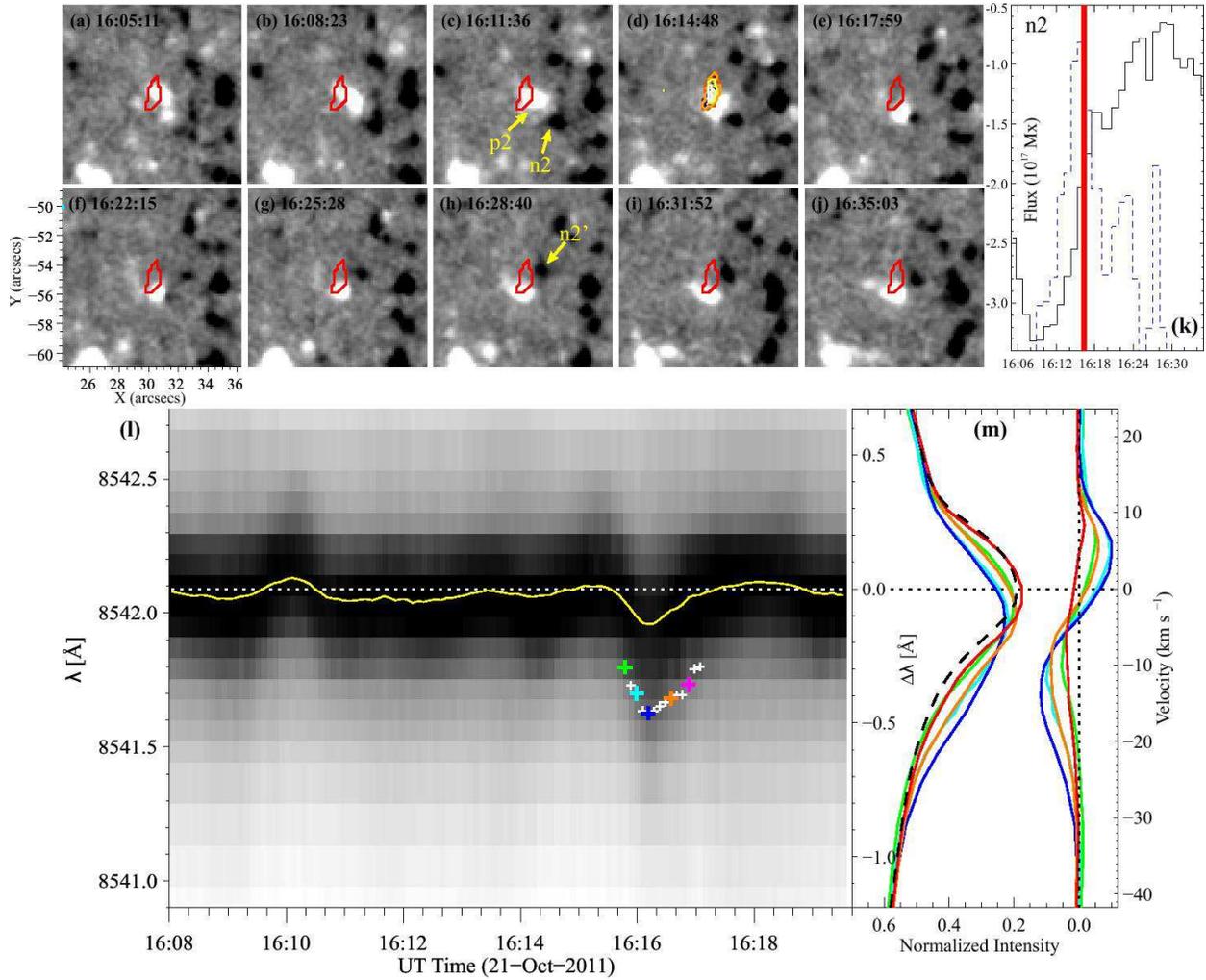}
  \caption{Same as Figure~\ref{f2} but for the RBE event 2 (indicated in Figure~\ref{f1}), which is associated with flux cancellation between p2 and n2. The blue dashed line in (k) is the time derivative of the magnetic flux of n2 (black line).} \label{f3}
\end{figure}

\begin{figure}[t]
  \epsscale{.6}
  \plotone{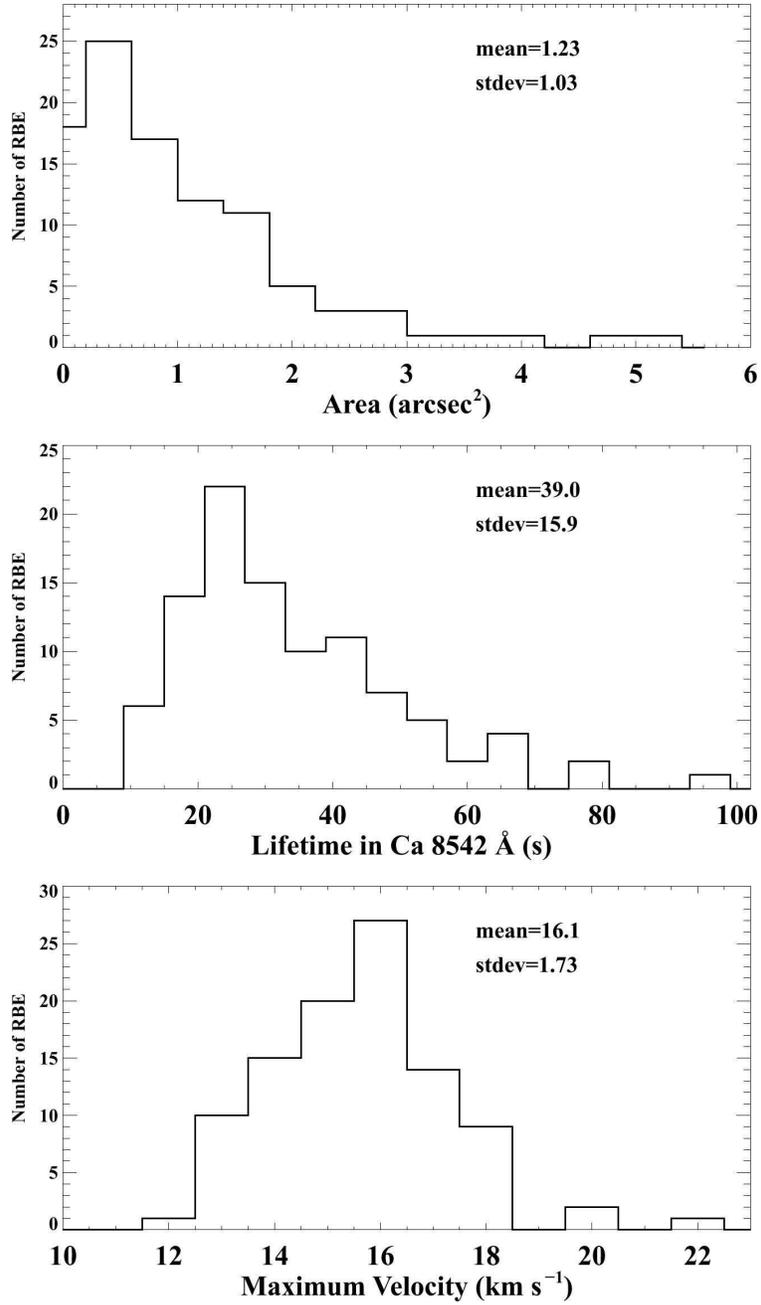}
  \caption{Histograms for area (top), lifetime in Ca~{\sc ii} 8542~\AA\ (b), and maximum Doppler velocity (bottom) of our observed RBEs.} \label{f35}
\end{figure}

To shed light on the driving mechanism of RBEs, we examine the associated photospheric magnetic field evolution using the high-resolution LOS magnetograms from \hinode/NFI. As an illustration, the RBE events 1 and 2 as labeled in Figure~\ref{f1} are presented as case studies. In the panels (a)--(i) of Figures~\ref{f2} and \ref{f3}, we show the time sequence of the underlying magnetic field, superimposed with the defined RBE region (red contour). The detected RBE features at selected instances are also overplotted in panel (d). For an easier evaluation, we choose the related magnetic element of negative polarity and plot the corresponding temporal evolution of magnetic flux in panel (k). We describe the events in detail as follows.

The RBE event 1 at the supergranule boundary is observed between about 16:17:04~UT and 16:17:38~UT. Right beneath the chromospheric RBE, a pair of magnetic elements p1-n1 of opposite polarity emerges from below the photosphere about seven minutes before the RBE occurrence (Figures~\ref{f2}(a)--(d)). The magnetic flux of the negative element n1 keeps increasing till 16:17~UT and then switches to a continuous decrease by \sm$10^{17}$~Mx (Figure~\ref{f2}(k)), as n1 seemingly cancels with the adjacent positive field p1' (Figures~\ref{f2}(e)--(h)). The transition from flux emergence to cancellation is simultaneous with the RBE (Figure~\ref{f2}(k)), which migrates slightly southward (Figure~\ref{f2}(d)). The structure and evolution of these magnetic elements suggest that this RBE may undergo a miniature version of the eruption as in the classical coronal jet model \citep[e.g.,][]{Shibata+etal2007Sci...318.1591S,liu11,Moore+etal2011ApJ...731L..18M}, where the emerging flux presses against the ambient opposite-polarity field to cause magnetic reconnection, driving RBEs. A consistent signature of energy release is the slight but discernible increase of the line core intensity cotemporal with the RBE occurrence (see Figure~\ref{f2}(m)). It is possible that the reconnection occurs at the low chromospheric level as the subsequent flux cancellation closely follows (see Section~\ref{sec:RBEwithCanEmg} for further discussion). The reduction of the p1 flux due to its cancellation with n1 is not obvious possibly due to the also emerging bipole p1'-n1'. It can be seen that n1' further cancels with a southern positive element from \sm16:28~UT (Figures~\ref{f2}(g)--(j)), which might lead to another episode of RBE activity at roughly the same location \citep{Sekse+etal2013ApJ...764..164S}.

Another RBE event 2 during about 16:15:59--16:16:40~UT is cospatial with the positive magnetic element p2 also located at the supergranule boundary (see Figure~\ref{f1}(b)). To the southwest lies the negative magnetic element n2 (Figures~\ref{f3}(a)--(b)), which is formed after coalescence of smaller magnetic patches \citep{Lamb+etal2013ApJ...774..127L}. Probably driven by convective flows, n2 moves toward and cancels with p2 starting from \sm16:12~UT (Figures~\ref{f3}(c)--(g)), until n2 dies out in 15 minutes with a total canceled flux of \sm2.5~$\times 10^{17}$~Mx. Interestingly, the cancellation rate peaks around 16:16~UT, cotemporal with the onset of the RBE (see Figure~\ref{f3}(k)). The close spatiotemporal correlation is tempting to infer an intrinsic relation between the RBE and the flux cancellation. This kind of cancellation may be a consequence of magnetic interaction between network and intranetwork fields, as the supergranular convective flows sweep the latter towards the former \citep[e.g.,][]{WangH+etal1998SoPh..178...55W,2014arXiv1408.2369G}. Later, p2 cancels again with another negative element n2' (Figures~\ref{f3}(h)--(j)), which could be associated with a recurrent RBE event.

\subsection{Statistical Association of RBEs with Magnetic Field Evolution}\label{sec:RBEwithCanEmg}

We attempt to explore whether the chromospheric RBEs are preferably associated with certain types of photospheric magnetic field evolution. To this end, we first resort to SWAMIS to automatically detect magnetic flux cancellation and emergence in our \hinode/NFI magnetic field data \citep{chen14}. We set the high threshold to 16.5~G (3$\sigma$) and the low threshold to 11~G (2$\sigma$), which SWAMIS uses to discriminate magnetic features based on the ``clumping'' algorithm. The recognized features are associated over time using a dual-maximum-overlap criterion, which can identify persistent features across image frames. For the best result, we added extra criteria that a feature must have an area of at least 4 pixels (0$\farcs$41$^2$) in each frame and its lifetime must be greater than \sm3 minutes (3 frames). The interaction among features is traced and classified into different types. A flux cancellation takes place when two adjacent (with a separation $\leqslant$5 pixels, or 0$\farcs$8) opposite-polarity features exhibit a flux decrease in at least 3 successive frames. The magnetic polarity inversion line between them is recorded as the cancellation site. A flux emergence is defined as a paired birth of opposite-polarity magnetic features, which are separated by at most 5 pixels. In the SWAMIS implementation, the requirement of flux conservation for the detected cancellation and emergence was set to be as permissive as physically reasonable --- the changes in fluxes of the detected features only had to agree in sign, not in magnitude.

Next we define a circular region of radius $r$ centered on each observed RBE, within which we search for the SWAMIS detected magnetic events during a time period $\tau$. Higher limits of $\tau \approx 1$~hr and $r \approx 2 \farcs 5$ are used based on the following justifications. (1) Usually the small-scale magnetic reconnection occurs above the photosphere, and the newly formed inverse U-loop would retract and submerge below the photosphere at a descent speed of about 0.3--1~\kms\ \citep{Zwaan1987ARA&A..25...83Z,harvey99}. This leads to a time delay between magnetic reconnection and the ensued canceling magnetic features, which were found to be tens of minutes depending on the height of the reconnection region \citep[e.g.,][]{harvey99,yurchyshyn01,chae10}. If magnetic loops submerge from an altitude of 1.2--1.5~Mm, which is the formation height of the \CaII\ 8542 spectral line \citep{mein80}, the time delay might be up to order of 1 hr. Here the implied possibility that RBEs could be driven by events of magnetic reconnection distributed over a range of heights was also suggested by \citet{Langangen+etal2008ApJ...679L.167L}. (2) Within 1~hr, a photospheric magnetic element can travel up to $2\farcs5$ given a velocity of \sm0.5~\kms\ \citep{giannattasio14}.

\begin{figure}[t]
  \epsscale{1.}
  \plotone{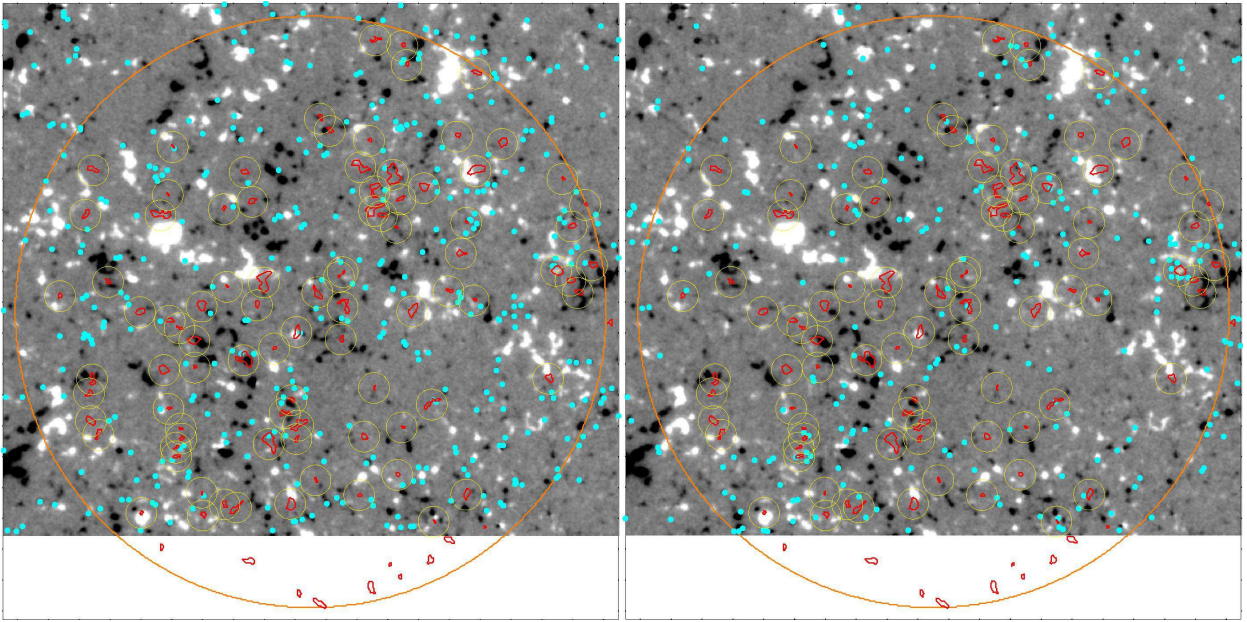}
  \caption{Same as Figure~\ref{f1}(b) but further overplotted with locations of the SWAMIS detected flux cancellation (cyan dots in (a)) and emergence (cyan dots in (b)). The cancellation events occurred during 16:08--17:08~UT and the emergence events during 15:20--16:20~UT. Each identified RBE is also surrounded by a circle (with a radius of 2$\farcs$5), representing the search region for the possibly associated magnetic events. See Section~\ref{sec:RBEwithCanEmg} for details.} \label{f4}
\end{figure}

As our 11 minute IBIS observation (16:08:02--16:19:31~UT) lies in the middle of the \sm2 hr \hinode/NFI data from 15:00--17:10~UT, we inspect magnetic cancellation events in the later half (16:08--17:08~UT) of the magnetogram sequence within the circular search region around each RBE. Similarly, flux emergence events are inspected in the early half (15:20--16:20) of the magnetic data. In Figures~\ref{f4}(a) and (b), we use cyan dots to mark the locations of the 378 cancellation events during 16:08--17:08~UT and those of the 246 emergence events during 15:20--16:20~UT as detected by SWAMIS, respectively. The 2$\farcs$5 radius search region is also denoted by the yellow circles around the RBEs (red contours). As a result, we find that among the 88 RBEs in the \hinode\ FOV, 43 (49\%) are associated with one or more cancellation; meanwhile, 34 (39\%) are associated with one or more emergence. There are 20 (23\%) RBEs that are associated with both flux emergence and cancellation, while 31 (35\%) RBEs are associated with neither of them.

\begin{figure}[t]
  \epsscale{.7}
  \plotone{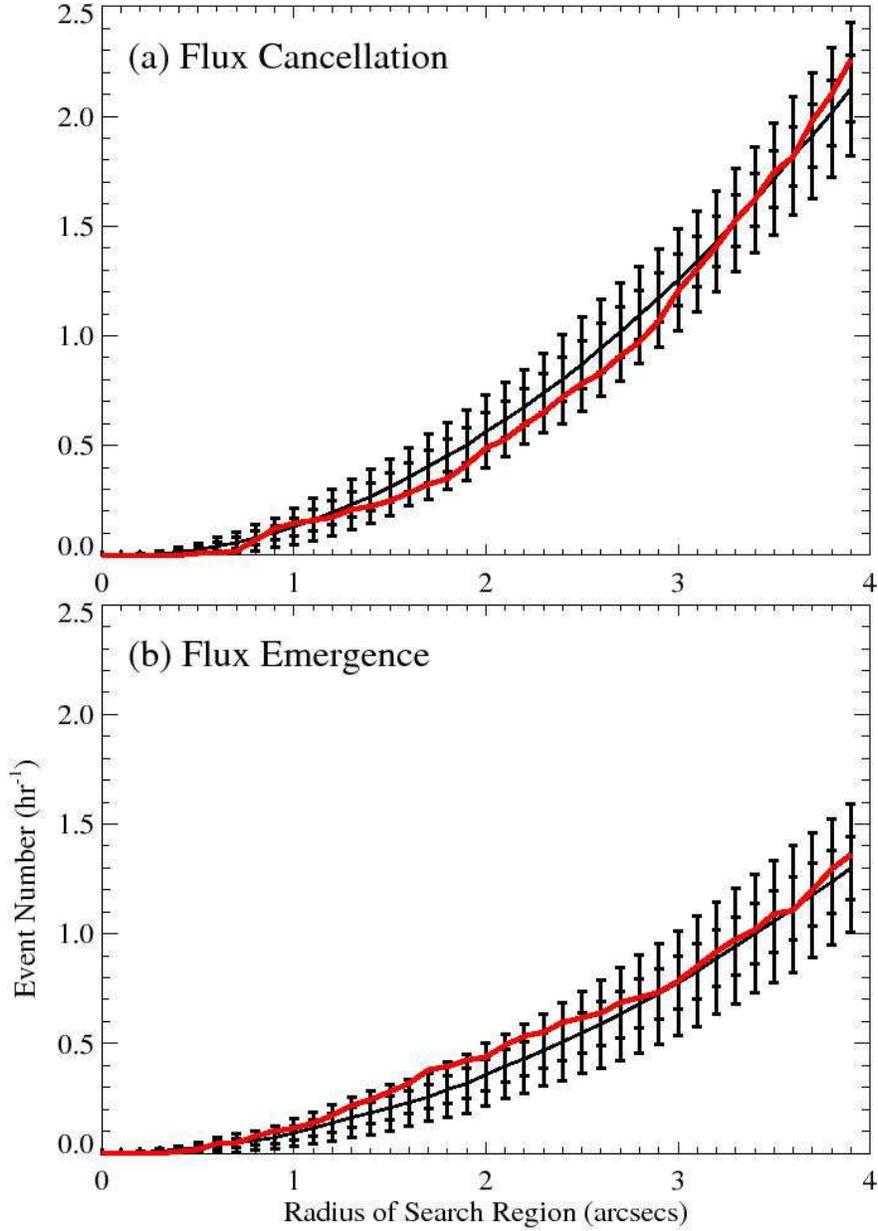}
  \caption{The radius of the circular search region versus the average number of magnetic flux cancellation (a) and emergence (b) events that can be found within (in a 1 hr period). The black lines with one and two sigma error bars are for the randomly selected locations in the \hinode\ FOV, and the red lines for the average occurrence counted around the RBE sites.} \label{f5}
\end{figure}

To evaluate the significance of these correlations, we randomly select 88 locations in the \hinode\ FOV and compute the average number of magnetic events that can be found within a search region of radius $r$. This randomization is repeated for 100 times for a specific $r$ and we also vary $r$ up to 4\arcsec. The same calculation is also made for the RBE locations. In Figure~\ref{f5}, we plot $r$ versus the average number of cancellation/emergence events that can be found, for both cases of the randomly selected locations (black lines with error bars) and the RBE sites (red lines). It appears that no matter what radius of the search region is used, the association of these RBEs with magnetic activities fall within two sigma of the random occurrence. As a further test, we repeat the evaluation after shortening the search time window $\tau$ to 40 and also 20 minutes. Still, the RBE-related magnetic cancellation (emergence) lies within two (three) sigma of the random occurrence. These results imply that the afore-described correlations may not be statistically significant; that is, these detected chromospheric RBEs are not statistically associated with photospheric magnetic events. Nonetheless, we caution that the lack of a statistical correlation might be partially attributed to the small data sample we analyzed (88 RBEs within 11 minutes), and also to the very quiet nature of this region. Obviously, more extended work needs to be done before definitive conclusions can be drawn.

\section{SUMMARY AND DISCUSSION}\label{sec:summary}
In this paper, we have taken advantage of the high spatiotemporal resolution data obtained by our campaign observation with IBIS and \hinode\ to study RBEs, the disk counterparts of type II spicules. The chromospheric RBE events are determined and characterized based on their spectral characteristics. The association of RBEs with photospheric magnetic field evolution is also explored using case studies as well as a statistical analysis aided by SWAMIS. Our results can be summarized as follows.

\begin{enumerate}

\item We identify a total of 98 \CaII\ 8542 RBEs in a quiet-Sun region of a 95\arcsec\ diameter round IBIS FOV at the disk center during 16:08:02--16:19:31~UT on 2011 October 21. Different from previous observations, this is a very quiet region free of obvious chromospheric rosette structure. The observed RBEs are either roundish or elongated (cf. \citealt{Langangen+etal2008ApJ...679L.167L}), and they are mainly found around the boundary of the supergranular cells, with some residing inside the cells.

\item These RBEs possess an average Doppler velocity of about 16~\kms, a mean lifetime in \CaII\ 8542~\AA\ of 39~s, and an average area of 1.2 arcsec$^2$. Their acceleration ranges from 200 to 700~\mss\ with a mean around 400~\mss. Notably, an additional deceleration phase ranging from $-$90 to $-$240~\mss\ with a mean around $-$160\mss\ is newly discovered to often closely follow the acceleration. These quantitative characterization of RBE properties are made possible by the detailed temporal evolution of spectra from the high-cadence IBIS spectroscopic observation. We nevertheless caution that some of the identified RBEs may include or modulated by acoustic shocks.

\item A clear example is presented where the RBE appears to be well associated with signatures of small-scale magnetic reconnection, i.e., the preceding magnetic flux emergence and the subsequent cancellation, which is in line with the classical reconnection-driven jet model \citep[e.g.,][]{Moore+etal2011ApJ...731L..18M}. In another sample, the RBE shows a close correlation with the magnetic cancellation, and could result from the intranetwork-network interaction sustained by the supergranular flows \citep[e.g.,][]{WangH+etal1998SoPh..178...55W}. Evidence from the magnetic field evolution also supports that RBEs may recur at similar locations \citep{Sekse+etal2013ApJ...764..164S}.

\item In order to explore the statistical association between chromospheric RBEs and photospheric magnetic field evolution, automatic magnetic events detection with SWAMIS is utilized. However, based on our limited RBE samples such an approach currently yields a result no better than random probability. This reflects the intrinsic nature that both the phenomena of RBEs and small-scale magnetic activities occur very frequently at similar locations, which poses a challenge for determining their correlation.

\end{enumerate}

Although not permitting definite conclusions, our results are suggestive to warrant further research. For instance, as our analyzed region is very quiet unlike previous RBE observations, future studies should be made on the association between RBEs and photospheric field evolution in quiet-Sun regions with clear magnetic field concentrations, as well as in other types of regions such as plages and coronal holes. Extended studies should also examine the relationship in a reverse way compared to the present work; that is, based on an observed flux cancellation (emergence) event, the probability of RBE occurrence within the preceding (following) \sm1 hr period needs to be investigated. An initial analysis based on our current data set shows that in a 20 minute time window (16:08--16:28~UT), there are total 108 cancellation events in the IBIS FOV, 83 of which are not within the 2.5\arcsec\ search region around the detected RBEs (in 16:08--16:19~UT). This may suggest that 77\% cancellation events (in 16:08-16:28 UT) do not have associated RBEs (in 16:08-16:19 UT). However, it is still possible that some related RBEs might occur out of our IBIS observation time period. Therefore, this kind of study requires good spectroscopic observation that has a sufficient temporal coverage.

\acknowledgments
IBIS is a project of INAF/OAA with additional contributions from University of Florence and Rome and NSO. NSO is operated by the Association of Universities for Research in Astronomy under a cooperative agreement with NSF, for the benefit of the astronomical community. \hinode\ is a Japanese mission developed and launched by ISAS/JAXA, with NAOJ as domestic partner and NASA and STFC (UK) as international partners. It is operated by these agencies in co-operation with ESA and NSC (Norway). We thank the anonymous referee for valuable comments. N.D., X.C., C.L., A.T., K.R., C.D., S.W., and H.W. were partially supported by NASA grants NNX11AO70G and NNX14AC12G. C.D. was also supported by grant DE 787/3-1 of the Deutsche Forschungsgemeinschaft (DFG). J.J. was partially supported by NASA grant NNX11AQ55G and NSF grant AGS 1250374. D.A.L. and C.E.D. were partially supported by NASA grants NNX08AJ06G and NNX11AP03G. R.L. was partially supported by NSF grant AGS 1153226. The research has made use of NASA's Astrophysics Data System.

\newpage

\end{document}